\documentclass[aps,prb,twocolumn,superscriptaddress,showpacs,showkeys]{revtex4-1}

\usepackage{amsfonts}
\usepackage{amssymb,amsmath}
\usepackage{graphicx} 
\usepackage{caption, subcaption}
\usepackage{bm}
\usepackage{tabularx}
\usepackage{color}
\newcolumntype{C}{>{\centering\arraybackslash}X}
\newcolumntype{M}[1]{>{\centering\arraybackslash}m{#1}}

\newcommand{\newc}{\newcommand}
\newc{\beq}{\begin{equation}}
\newc{\eeq}{\end{equation}}
\newc{\beqa}{\begin{eqnarray}}
\newc{\eeqa}{\end{eqnarray}}
\newc{\mbf}{\mathbf}
\newc{\dg}{\dagger}
\newc{\rgtaw}{\rightarrow}
\newc{\lftaw}{\leftarrow}
\newc{\lrgtaw}{\longrightarrow}
\newc{\llfttaw}{\longrightarrow}

\begin{document}

\title{The spin-1/2 Heisenberg antiferromagnet on the pyrochlore lattice: An exact diagonalisation study}

\author{V. Ravi Chandra}
\email{ravi@niser.ac.in}
\affiliation{School of Physical Sciences, National Institute of Science Education and Research Bhubaneswar, HBNI, Jatni, Odisha 752050, India}
\author{Jyotisman Sahoo}
\email{jsahoo@iastate.edu}
\affiliation{Department of Physics and Astronomy, Iowa State University, Ames, Iowa 50011, USA}


\begin{abstract}
We present exact diagonalisation calculations for the spin-1/2 nearest neighbor antiferromagnet on the pyrochlore lattice.
We study a section of the lattice in the $[111]$ direction and analyse the Hamiltonian of the breathing pyrochlore system with two coupling
constants $J_1$ and $J_2$ for tetrahedra of different orientations 
and investigate the evolution of the system from the limit of disconnected tetrahedra ($J_2 = 0$) to a correlated state at $J_1=J_2$. 
We evaluate the low energy spectrum, two and four spin correlations, and spin chirality correlations
for a system size of up to 36 sites. The model shows a fast decay of spin correlations 
and we confirm the presence of several singlet excitations below the lowest magnetic excitation.  
We find chirality correlations near $J_1=J_2$ to be small at the length scales available at this system size.  
Evaluation of dimer-dimer correlations and analysis of the nature of the entanglement of the tetrahedral unit shows that the triplet sector of the tetrahedron contributes significantly to
the ground state entanglement at $J_1 = J_2$.
\end{abstract}

\maketitle

\section{Introduction}
The pyrochlore lattice is one of the earliest lattices to be investigated in the study of geometrically frustrated
magnetic systems. Shortly after foundational work on the triangular lattice \cite{wannier_triangular_lattice_1950}, 
the pyrochlore antiferromagnet was among the first three dimensional models for which it was established
that the nearest neighbour exchange does not result in magnetic ordering and that the ground 
state has a finite entropy \cite{anderson_Ising_pyrochlore_1956}: properties which are
now routinely used to characterise the extent of frustration in a magnetic system. The interest in the physics of 
this lattice has continued unabated over several decades. Apart from intrinsic theoretical
interest this is also because of the existence of real materials with this crystal structure. Examples are the spinels $A B_2 O_4$ and  
the $A_2 B_2 O_7$ oxides \cite{gardner_gingras_greedan_rmp} where the magnetic ions reside on the pyrochlore lattice.
Interestingly, among the Hamiltonians that have been used to model various classes of materials with this general structure 
the simple spin-1/2 nearest neighbour Heisenberg antiferromagnet has received relatively less attention in the past. The last few years
has seen a renewed interest in this system which in part is thanks to the discovery of new materials which can be modeled
as spin-1/2 Heisenberg antiferromagnets \cite{breathing_pyrochlore_kimura_et_al_2014, molybdate_pyrochlore_gaulin_et_al_2014, Yiqbaletal_gearwheel_spin_liquid}. \\

Many techniques have been used to study this model 
and like several other magnetic systems with geometric frustration or competing 
interactions there is no clear consensus on the nature of the low temperature phase. The ground
state prediction for the Hamiltonian varies with the technique used to study the problem. The ground state manifold of  
a single tetrahedron of four spins, the basic unit of the lattice, consists of two degenerate singlet states. Approaches 
which restrict the Hilbert space of tetrahedra to the subspace of these singlets and derive effective theories for the
Hamiltonian usually result in a dimer singlet phase. This phase with broken translational symmetry has long range 
order in the dimer-dimer correlations 
\cite{harris_bruder_berlinsky, tsunetsugu_2001_pyrochlore_prb,tsunetsugu_pyrochlore_jpsj, isoda_mori_pyrochlore_jpsj, tsunetsugu_2017_pyrochlore_ptep, berg_altman_auerbach_prl}.
A power series expansion of the density matrix of the problem in powers of the inter-tetrahedra coupling
results in a spin liquid \cite{canals_and_lacroix_pyrochlore_prl} with a very short correlation length for two spin correlations.
The studies based on the analysis of Sp(N) models support a (spontaneously) broken inversion symmetry \cite{goerbig_moessner_sondhi_qdm_prb} 
or the presence of several different saddle points at large N \cite{tchernyshyov_moessner_sondhi_large_N_epl} but remain noncommittal about 
the SU(2) limit of those Hamiltonians.
Fermionic mean field theory followed by a variational Monte Carlo analysis of a mean field state lends support
to a chiral spin liquid \cite{burnell_et_al_pyrochlore_prb, kim_and_han_pyrochlore_prb}  with long range order in the scalar chirality operator 
$\mathbf{S}_i \cdot (\mathbf{S}_j \times \mathbf{S}_k)$, where $(i,j,k)$ are  
on a tetrahedron. Finally, a finite temperature analysis using diagrammatic Monte Carlo method predicts spin
ice correlations even in the Heisenberg limit \cite{svistunov_et_al_dmc_prl}, albeit at finite temperatures. 

\begin{figure*}[t]
\centering
\begin{subfigure}[h]{0.45\textwidth}
    \centering
    \includegraphics[scale=0.68]{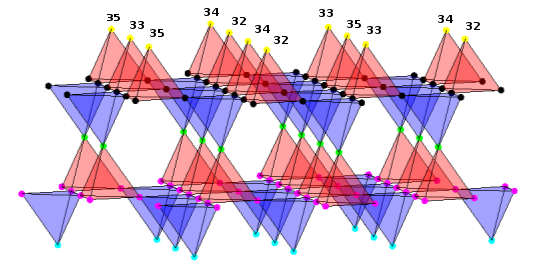}
\vspace{0.5cm}
    \subcaption{}
    \label{Fig_111_pyro}
\end{subfigure}%
~
\begin{subfigure}[h]{0.5\textwidth}
    \centering
    \includegraphics[scale=0.5]{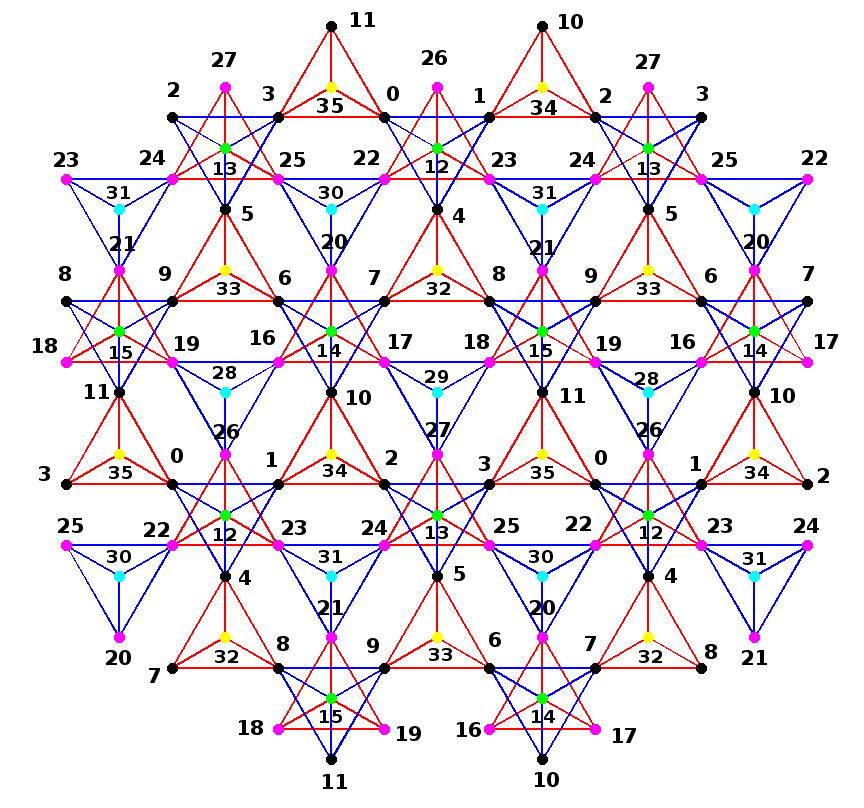}
    \subcaption{}
    \label{Fig_111_pyro_2d}
\end{subfigure}
\caption{{{\small{The pyrochlore lattice: The figure on the right is a 2D projection with the [111] axis out of the page. The black and magenta
dots are the two Kagome layers and the yellow, green and the cyan dots are the top, middle and bottom triangular lattice layers respectively. The numbers in the
figure on the right is the site indexing used throughout the paper. The blue lines are the $J_1$ bonds and the red lines are the $J_2$ bonds. 
The top layer site labels are also presented in the 3D figure to establish the orientation.}}}}
\label{Fig_pyro}
\end{figure*}

The focus of this paper is the analysis of the spin-1/2 nearest neighbour antiferromagnet on the pyrochlore lattice using exact 
diagonalisation (ED). The principal and obvious limitation of ED studies is that we usually cannot do meaningful finite size 
scaling for many systems of current interest, especially if they are in two or three dimensions
and the ground state is (magnetically) disordered but correlated. However, the technique does offer the advantage that 
for the system sizes that can be handled using this technique it provides essentially exact information about the lattices 
studied. We present here the results of such an ED study for the nearest neighbour spin-1/2 antiferromagnetic pyrochlore lattice for a 
system size of up to $36$ sites. We focus on a section of the lattice in the [111] direction and evaluate energies, two, four and six point correlators. In Sec. \ref{results_sec_lat_ham}
we introduce the Hamiltonian being studied and details of the finite size section being investigated. In Sec. \ref{section_results_spectrum} we discuss the nature
of the spectrum in the vicinity of the ground state. In Sec. \ref{gscorrs} 
we present all the evaluated correlations and the analysis of the ground state based on these correlations.    

\section{Lattice and the Hamiltonian}
\label{results_sec_lat_ham}

The pyrochlore lattice is a lattice of corner sharing tetrahedra which can be seen as a face centered
cubic lattice with a four site basis (see Fig. \ref{Fig_pyro}).
Using the conventional FCC primitive lattice vectors
${\mbf{a_1}}=\left(\frac{1}{2}, \frac{1}{2}, 0 \right ),{\mbf{a_2}} = \left(0, \frac{1}{2}, \frac{1}{2} \right), {\mbf{a_3}}=\left(\frac{1}{2}, 0, \frac{1}{2} \right)$
the four sublattices are: $\left (0, 0,0 \right), \frac{{\mbf{a_1}}}{2},  \frac{{\mbf{a_2}}}{2},  \frac{{\mbf{a_3}}}{2}$.
Each site is connected to six nearest neighbours and is part of two tetrahedra of different orientation as can be seen in Fig. \ref{Fig_pyro}. \\

The model Hamiltonian being studied in this {\mbox{paper is :}}
\beq
H = J_1 \sum_{<ij>,A} \mathbf{S}_{i} \cdot \mathbf{S}_j + J_2 \sum_{<ij>, B} \mathbf{S}_{i} \cdot \mathbf{S}_j
\label{ham_def}
\eeq
$\mathbf{S}_{i}$ are spin-1/2 operators at the sites of the pyrochlore lattice.
$J_i$ are nearest neighbour antiferromagnetic coupling constants for the tetrahedra of two different 
orientations ($J_1$ for blue and $J_2$ for red tetrahedra in Fig. \ref{Fig_pyro}). The parameter in the problem 
is $J_2$ ($J_1$ is set to 1 and so is $\hbar$). This Hamiltonian is called the breathing pyrochlore Hamiltonian \cite{breathing_pyrochlore_kimura_et_al_2014}.\\

\begin{figure*}[t]
\centering
\begin{subfigure}[h]{0.5\textwidth}
    \centering
    \includegraphics[scale=0.5]{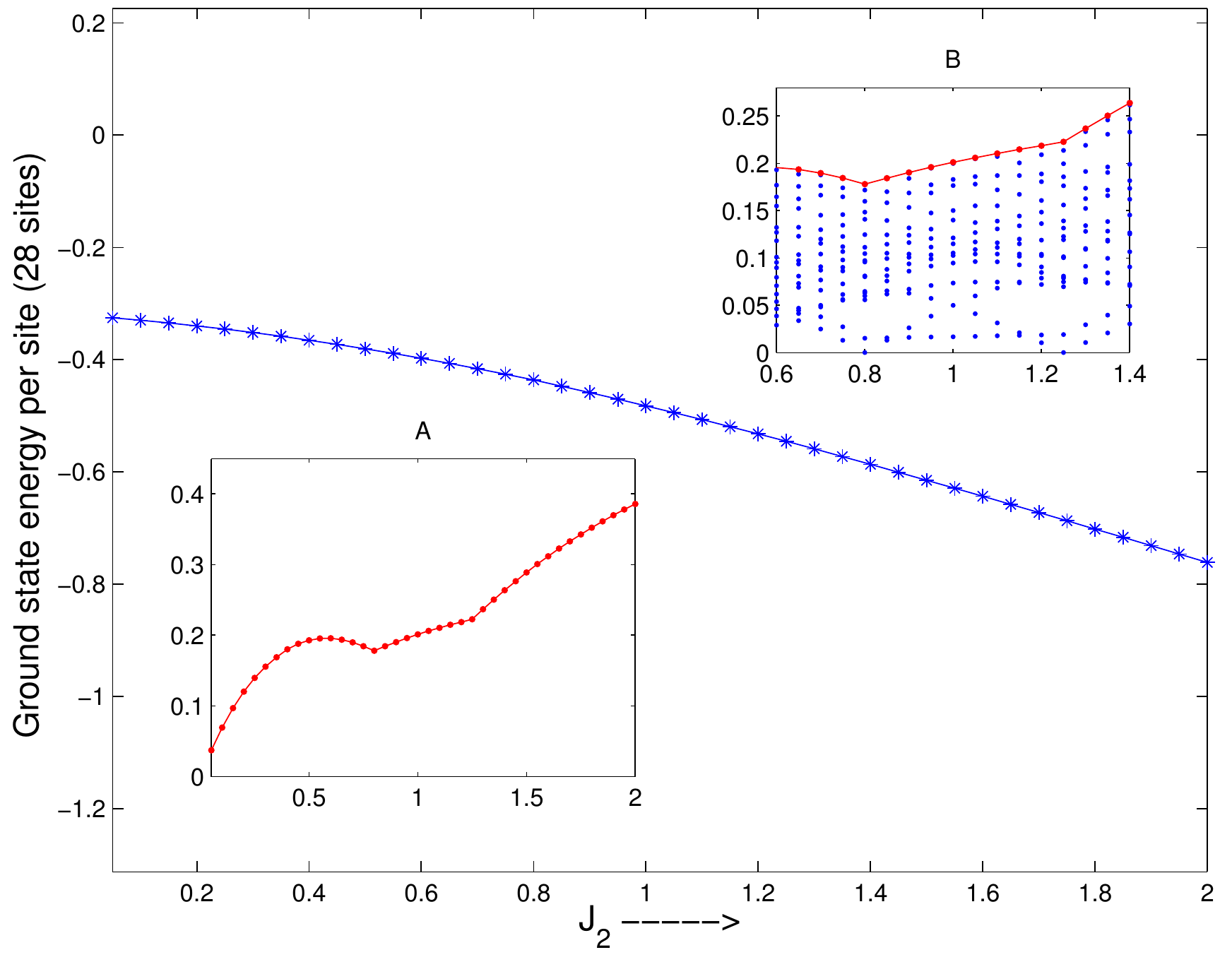}
    \label{Fig_spectrum_28_sites}
\end{subfigure}%
~
\begin{subfigure}[h]{0.5\textwidth}
    \centering
    \includegraphics[scale=0.5]{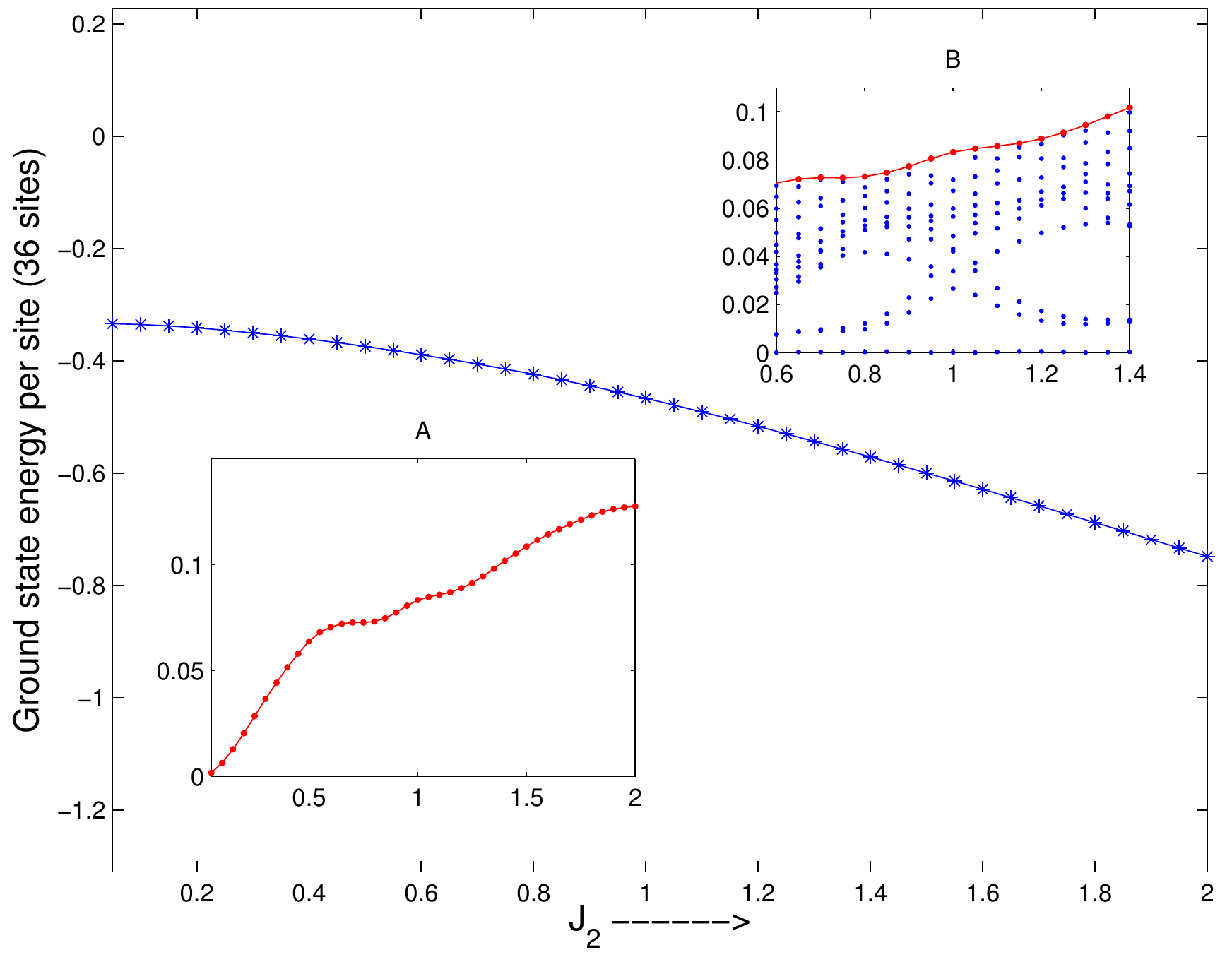}
    \label{Fig_spectrum_36_sites}
\end{subfigure}
\caption{{\small{The ground state energy per site and the nature of the low energy spectrum as a function of $J_2$. 
The left figure is for the $28$ site cluster and the right figure for the $36$ site cluster. The inset labelled $A$ 
shows the spin gap (the energy difference between the lowest magnetic excitation and the singlet ground state) and 
the inset labelled $B$ shows the presence of singlet excitations below the spin gap (blue markers) near $J_2 = 1$. 
Energy eigenvalues plotted are for the Hamiltonian in Eq. \ref{ham_def} with $J_1$ and $\hbar$ 
set to 1. See section \ref{section_results_spectrum} for details.}}}
\label{Fig_spectrum}
\end{figure*}

We study a finite part of the lattice shown in Fig. \ref{Fig_111_pyro} using exact diagonalisation.  
Fig. {\ref{Fig_111_pyro_2d}} shows a 2D projection of the system studied with site labels. The data for two  
system sizes of $28$ and $36$ sites have been analysed in detail in this paper. The lattice of $28$ sites is obtained by omitting 
sites in the bottom and top triangular layers. Periodic boundary conditions are imposed in the Kagome and 
triangular lattice planes. The site indexing used to label the different sites shown in Fig. {\ref{Fig_111_pyro_2d}} 
will be used in the rest of the paper. 

We employ the conservation of the $S_z^{total} =  \sum_{i} S_{iz}$ and the spin inversion symmetry in the $S_z^{total} = 0$ 
sector to reduce the size of the Hilbert space. Using these two symmetries the Hilbert space dimension of the $36$ site system 
is $4537567650$. The Lanczos algorithm \cite{cullum_and_willoughby_lanczos_1985} is used throughout for computations of the ground state
and low lying excitations.
$J_2$ is varied from $0.0$ to $2$ and is the parameter in the problem and corresponds to the x-axis 
(which starts at $J_2=0.05$) in all plots unless mentioned otherwise. We note that the $J_2>1$ region is of course physically 
the same as a section of the $J_2<1$ region since for finite size systems considered here it simply corresponds to exchanging 
the roles of the two kinds of tetrahedra. Nevertheless we present data for the whole range mentioned as an additional explicit 
consistency check. There are several $J_2$ values for which using the above mentioned two symmetries still result in doubly 
degenerate lowest eigenvalues in the symmetry sector containing the ground state. This degeneracy stems from the symmetry of 
the Hamiltonian under certain permutations of site indices which are elements of the automorphism group for this graph. We 
choose one such permutation to extract orthogonal states with distinct symmetry labels for $J_2$ values where degeneracy of 
the ground state is present. The used permutation of site labels for the results presented in this paper is detailed in 
Appendix \ref{appendix_permutation}, and in the rest of the text is denoted by $\mathcal P$. The same symbol is used to 
denote the operator for that permutation of site indices or the eigenvalues of that operator where necessary.  We have 
verified that the computed energy eigenvalue and the energy expectation value that can be derived using the computed correlations 
match to at least $10^{-6}$. All explicitly stated numbers have been quoted with 6 significant digits, though several numbers are 
much more accurate.

At this point it is pertinent to recall some elementary properties of the basic tetrahedral unit which play a 
role in the analysis of the evaluated correlation functions.
A single tetrahedron with all equal strength antiferromagnetic bonds has as its ground state two degenerate singlets. This follows
directly from the fact that the Hamiltonian of the tetrahedron itself is simply equal to $\displaystyle{H_{tetrahedron} = \frac{J}{2}{(\sum_i \mbf{S_i})}^2  - \frac{3J}{2}}$.
There are distinct ways of thinking about this space of degenerate ground states which leads to consideration
of different kinds of ordered states on the full lattice. One obvious way is to group the four spins into distinct pairs of spins (say $[12][34])$).
Then we can build total spin singlets by either considering a direct product of singlets of each pair or by constructing a spin-0
state using the triplet states of each pair. The dimer-dimer correlation $\langle (\mbf{S}_1 \cdot \mbf{S}_2 ) (\mbf{S}_3 \cdot \mbf{S}_4) \rangle$ 
for the former is $9/16$ and $1/16$ for the latter.  
However, one could equally consider the Hamiltonian of the tetrahedron to be a sum of Hamiltonians defined on the triangular faces of the
tetrahedron : $H = \frac{1}{2} \sum_{\triangle} H_{\triangle} = \frac{1}{2} \sum_{ijk} \left ( \mbf{S}_i \cdot \mbf{S}_j +  \mbf{S}_j \cdot \mbf{S}_k  +  \mbf{S}_k \cdot \mbf{S}_i \right ) $, where
the sum is over all distinct triads of spins $(ijk)$ on the tetrahedron. It is well known \cite{wenwilczekzeechiral} that $\Xi_{ijk} =   \mbf{S}_i \cdot ( \mbf{S}_j \times  \mbf{S}_k)$
commutes with $\sum_i \mbf{S}_i$ and thence the eigenstates of $H_{tetrahedron}$ can be expressed as eigenstates of $\Xi_{ijk}$ belonging
to a face of a tetrahedron. Furthermore, if we focus on only the two dimensional ground state singlet manifold then suitable linear combinations
which are simultaneous eigenstates of all $\Xi_{ijk}$ can be constructed \cite{wenwilczekzeechiral}.
Also,  $(\Xi_{ijk})^2 = \frac{3}{32}  - \frac{1}{8} H_{\triangle}$, for spin-1/2. Thus the full Hamiltonian on the lattice can be written either as a sum
of total spin operators on tetrahedra or the sum of the squares of $\Xi_{ijk}$ operators on all triangles of the lattice.
Given this structure of the ground state manifold of the basic tetrahedral unit and the Hamiltonian, Eq. \ref{ham_def}
appears to be a natural candidate (especially at $J_1 = J_2$) to explore spontaneous dimerisation 
\cite{harris_bruder_berlinsky, tsunetsugu_2001_pyrochlore_prb,tsunetsugu_pyrochlore_jpsj, isoda_mori_pyrochlore_jpsj, tsunetsugu_2017_pyrochlore_ptep, berg_altman_auerbach_prl}
or long range chiral order in three dimensions \cite{burnell_et_al_pyrochlore_prb, kim_and_han_pyrochlore_prb}.  
We will analyse those possibilities from the point of view of exact diagonalisation calculations in the following sections. 

\begin{table*}[t]
  \centering
  \begin{tabularx}{0.9 \textwidth}{|M{7.0cm}|C|C|}
\hline
\hline
Property & 28 sites  & 36 sites \\
\hline
\hline
Ground state energy per spin  & $-0.482081$  & $-0.466971$ \\
\hline
Spin gap &  $0.201229$  &  $0.083273$ \\
\hline
$(\langle \mbf{S}_o \cdot \mbf{S}_i \rangle)_{1-bond}$ & -0.192600 & -0.168562\\
$(\langle \mbf{S}_o \cdot \mbf{S}_i \rangle)_{2-bond}$ & 0.035057 &  0.028212\\
$(\langle \mbf{S}_o \cdot \mbf{S}_i \rangle)_{3-bond}$ & -0.014860 & -0.008655\\
\hline
$\langle \phi_{0-12}~\phi_{1-4}\rangle, \langle \phi_{0-1}~\phi_{4-12}\rangle, \langle \phi_{0-4}~\phi_{1-12}\rangle$   &  0.152902  & 0.093991  \\
$\langle \phi_{0-12}~\phi_{6-14} \rangle $ & 0.074815 & 0.046312 \\
$\langle \phi_{0-12}~\phi_{2-13} \rangle$ & 0.016658  &  0.016217 \\
$\langle \phi_{0-12}~\phi_{23-31} \rangle $ &  & 0.040566 \\
$\langle \phi_{12-26}~\phi_{22-23} \rangle, \langle \phi_{12-23}~\phi_{22-26} \rangle, \langle \phi_{12-22}~\phi_{23-26} \rangle$  &  0.152902   & 0.093991  \\
$\langle \phi_{12-26}~\phi_{13-27} \rangle$ & 0.074815  &  0.046312 \\
$\langle \phi_{12-26}~\phi_{14-20} \rangle$ &  0.016657  &  0.016217  \\
$\langle \phi_{12-26}~\phi_{10-34} \rangle$ &    & 0.038405   \\
\hline
$\langle \Xi_{1-4-12} \Xi_{1-4-12} \rangle$ & 0.159053  &  0.151897 \\
$\langle \Xi_{12-23-26} \Xi_{12-23-26} \rangle$ &  0.159053  & 0.151897 \\
$\langle \Xi_{4-8-32} \Xi_{4-8-32} \rangle$ &   &  0.171852 \\
$\langle \Xi_{21-24-31} \Xi_{21-24-31} \rangle$ & & 0.171852 \\
$\langle \Xi_{1-4-12} \Xi_{7-10-14} \rangle$ & 0.020505  &  0.030824 \\
$\langle \Xi_{1-4-12} \Xi_{3-5-13} \rangle$ & 0.016554 &  0.006981 \\
$\langle \Xi_{1-4-12} \Xi_{21-24-31} \rangle$ & & -0.000230 \\
$\langle \Xi_{1-4-12} \Xi_{18-27-29} \rangle$ & &  0.000141 \\
\hline
\hline
\end{tabularx}
\caption{{{\small{{Spectrum and correlations in the ground state for $J_2=1$. $\phi_{i-j}$ denotes $\mbf{S}_i \cdot \mbf{S}_j $ and $\Xi_{i-j-k}$ denotes 
$\mbf{S}_i \cdot (\mbf{S}_j \times \mbf{S}_k) $. $\langle \rangle$ denotes the expectation value in the ground state. 
The site indices follow the indexing shown in Fig. \ref{Fig_111_pyro_2d}. Missing numbers in the second column correspond to correlations which contain sites
not present in the 28 site system. }}}}}
\label{j2onedatatable}
\end{table*}

\section{Low energy Spectrum}
\label{section_results_spectrum}

We begin with a description of the low energy spectrum of the system for the two lattice sizes considered.
Fig. \ref{Fig_spectrum} shows the variation of the ground state energy per spin as a function of $J_2$.
The values of the energies for two sizes at the point $J_2=J_1=1$ are $-0.482081$ (28 sites) and $-0.466971$ (36 sites) respectively.
The difference in the energy per site values of these two lattice sizes for other $J_2$ values is of similar magnitude or smaller. 

The insets in the figures show some details of the excitation spectrum above the ground state. Inset A shows 
the variation of the spin gap with $J_2$. Inset B depicts the  
low energy excitations below the lowest non-singlet state in a region close to $J_2=1$. The ground state
and the lowest non-magnetic excitation, which is the lowest energy state in a different symmetry sector, 
are fully converged, in the Lanczos sense. 
The singlets within that gap are partially converged to various degrees of accuracy at that Lanczos iteration. 
However the changes of their energies with iterations indicate that almost all of them are expected to be in 
the gap region after convergence to high accuracy. The spin gap for the 
28 and 36 site systems are $0.201229$ and $0.083273$ respectively for $J_2 = 1$.\\
As can be seen for both the lattice sizes there are several singlet excitations below 
the lowest energy magnetic state. This is a feature that has been seen in ED studies of other frustrated magnetic systems 
\cite{waldtmann_et_al_kagome_1998, palmer_and_chalker_checkerboard_ED_2001, fouet_et_al_checkerboard_2003}, 
most well known among them being the Kagome antiferromagnet. We would like to note that the density of singlets below the magnetic
gap depends on the boundary conditions imposed. For example, whereas about $180$ states below the lowest magnetic states are known
for the 36 site system of the Kagome lattice with periodic boundary conditions \cite{waldtmann_et_al_kagome_1998}, the number drops down 
to only a few for the same system with open boundary conditions. In our system we have periodic boundary conditions in two directions.
The singlet gap for the 28 site system is $0.016832$ whereas for the 36 site system it is $0.000143$ for $J_2=1$.
It is not immediately clear if the rather small size of the singlet gap for the $36$ site system is an artefact of our specific finite size geometry.
We have however explicitly verified that it is indeed a genuine excitation with the computed energy and not an artefact of the loss of orthogonality 
affecting the Lanczos vectors \cite{cullum_and_willoughby_lanczos_1985}. We cannot predict the fate of this 
small gap at larger length scales. We now present various spin correlations in the system to obtain a more detailed characterisation of the ground state.\\

\begin{figure*}[t]
\includegraphics[scale=1.0]{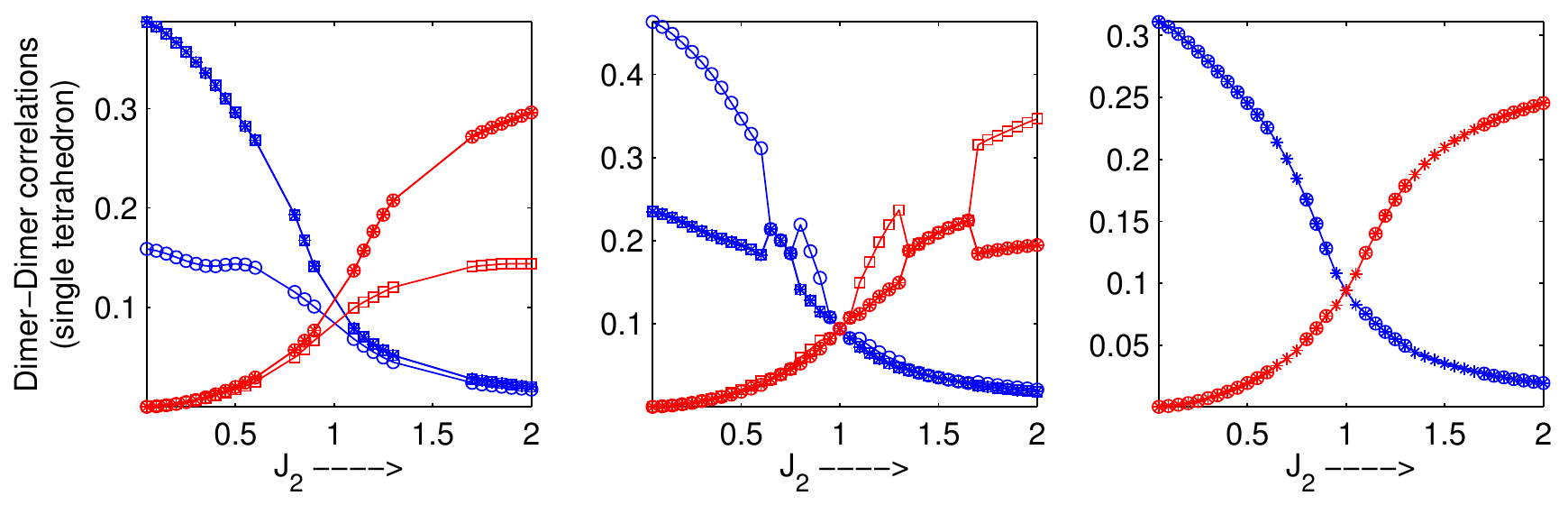}
\caption{{\small{{The variation with $J_2$ of dimer-dimer correlations within a single tetrahedron for the $36$ site cluster. Left figure and middle figures are for eigenvalue
$+1$ and $-1$ respectively of the site permutation operator $\mathcal{P}$ detailed in Appendix \ref{appendix_permutation}. The markers indicate the
following correlations in the ground state ( ${\phi}_{ij} \equiv  \mbf{S}_i \cdot \mbf{S}_j$) : blue stars $ \rgtaw \langle \phi_{0-1}~\phi_{4-12}\rangle$,
blue circles $  \rgtaw \langle \phi_{0-4}~\phi_{1-12}\rangle$, blue squares $ \rgtaw  \langle \phi_{1-4}~\phi_{0-12}\rangle$,  red stars
$\rgtaw \langle \phi_{12-26}~\phi_{22-23}\rangle$, red circles $\rgtaw \langle \phi_{12-23}~\phi_{22-26}\rangle$,
red squares $ \rgtaw  \langle \phi_{12-22}~\phi_{23-26}\rangle$. The figure on the right shows the average of all the three correlations
within a tetrahedron. The blue markers are for the $J_1$ tetrahedron red for the $J_2$ tetrahedron as in the other two figures. The open
circles are for eigenvalue $+1$ and the stars are for eigenvalue $-1$ of $\mathcal{P}$.}}}}
\label{Fig_dim_dim_single_tetrahedron}
\end{figure*}
\section{Ground state correlations}
\label{gscorrs} 
We evaluated the spin-spin ($\langle \mbf{S}_i \cdot \mbf{S}_j \rangle$), dimer-dimer ($\langle {\phi}_{ij}\cdot {\phi}_{kl} \rangle$, $ {\phi}_{ij} \equiv  \mbf{S}_i \cdot \mbf{S}_j$), 
and scalar spin chirality correlations ($\langle {\Xi}_{ijk}\cdot {\Xi}_{lmn} \rangle$, $\Xi_{ijk} \equiv \mbf{S}_i \cdot ( \mbf{S}_j \times  \mbf{S}_k)$ ) 
for the ground state of the Hamiltonian as a function of $J_2$. We discuss below in turn these three correlations to 
analyse the zero temperature phase of the system.
Table \ref{j2onedatatable} shows detailed information about the spectrum and spin correlations for $J_2=1$ and we will refer to it frequently in the following. 
As described in the introduction, this is the point in the phase diagram which has received most attention and has been conjectured to harbour several different kinds of ground states. 
The ground state for this point is a non-degenerate singlet which belongs to the eigenvalue $-1$ for the site permutation operator $\mathcal P$. 
We note that the 28 site system has several incomplete tetrahedra whereas all sites of the 36 site system are
part of (at least one) tetrahedron which has all six bonds. To that extent the correlations of latter system are expected to have
properties more aligned with the full lattice. For this reason unless mentioned otherwise the plots in this section are all for the system with 36 sites.

\subsection{Two spin correlations}
\label{two_spin_corrs_section}
We begin with a brief description the behaviour of the two spin correlations $\langle \mbf{S}_o \cdot \mbf{S}_i \rangle$. We evaluate these correlations
using the site with index $0$ as the reference site (see Fig. \ref{Fig_111_pyro_2d}). For $J_2 = 1$ in Table. \ref{j2onedatatable}
 $(\langle \mbf{S}_o \cdot \mbf{S}_i \rangle)_{1-bond}$ means the average of correlations with all sites which can be reached from site $0$ 
by crossing one bond. Analogous meanings follow for the other two correlations.
As can be seen very clearly the correlations decay rapidly with distance. This rapid decay is consistent with an expectation of a magnetically disordered 
ground state in this highly frustrated system and has been reported in earlier analyses \cite{canals_and_lacroix_pyrochlore_prl}. 
Choosing another site as a reference site might result for this finite size system in (quantitatively) 
different behaviour since translational invariance is absent in the third direction.  However, we
have verified that the general trend of rapid decay holds and does so for all values of $J_2$. We will not present here further
details of two point correlations as the general feature that correlations weaken significantly beyond the smallest length
scale is true for all $J_2$. We present in the following sections higher order spin correlations.

\subsection{Dimer Dimer correlations}
\label{dimer_dimer_corrs}
We now study the dimer-dimer correlations in the ground state.
They are denoted in the paper by $\langle \phi_{i-j}~\phi_{k-l} \rangle$ where $\phi_{i-j} \equiv \mbf{S}_i \cdot \mbf{S}_j $.\\ 

In the table \ref{j2onedatatable} we show some relevant dimer-dimer correlations for $J_2 = 1$.   
The table shows two different kinds of dimer-dimer correlations. The first are ($\langle \phi_{0-12}~\phi_{1-4}\rangle, \langle \phi_{0-1}~\phi_{4-12}\rangle, \langle \phi_{0-4}~\phi_{1-12}\rangle$)
and ($\langle \phi_{12-26}~\phi_{22-23} \rangle, \langle \phi_{12-23}~\phi_{22-26} \rangle, \langle \phi_{12-22}~\phi_{23-26} \rangle$)
which are the three possible dimer-dimer correlations within a tetrahedron, the former being a $J_1$ tetrahedron and the latter a $J_2$ tetrahedron. At $J_2=1$, all the
three different dimer-dimer correlations have the same value. 

\begin{table}[h]
\centering
\begin{tabularx}{0.5 \textwidth}{|C|C|}
\hline
$J_2$ range & Eigenvalue of $\mathcal{P}$ \\
\hline
$(0, 0.6)$ & $(+1, -1)$ \\
$(0.65, 0.75)$ & $(-1)$ \\
$(0.8, 0.9)$ & $(+1, -1)$\\
$(0.95, 1.05)$ & $(-1)$ \\
$(1.1, 1.3)$ & $(+1, -1)$\\
$(1.35, 1.65)$ & $(-1)$ \\
$(> 1.65)$ & $(+1, -1)$\\
\hline
\end{tabularx}
\caption{{\small{{The different regions of $J_2$ and eigenvalue of the symmetry operator $\mathcal{P}$ for the ground state of the $36$ site cluster. Entries with only one eigenvalue for $\mathcal{P}$ correspond
to a non-degenerate ground state. (Note: We vary $J_2$ in increments of $0.05$) }}}}
\label{table_deg_regions}
\end{table}

\begin{figure*}[t]
\includegraphics[scale=1.0]{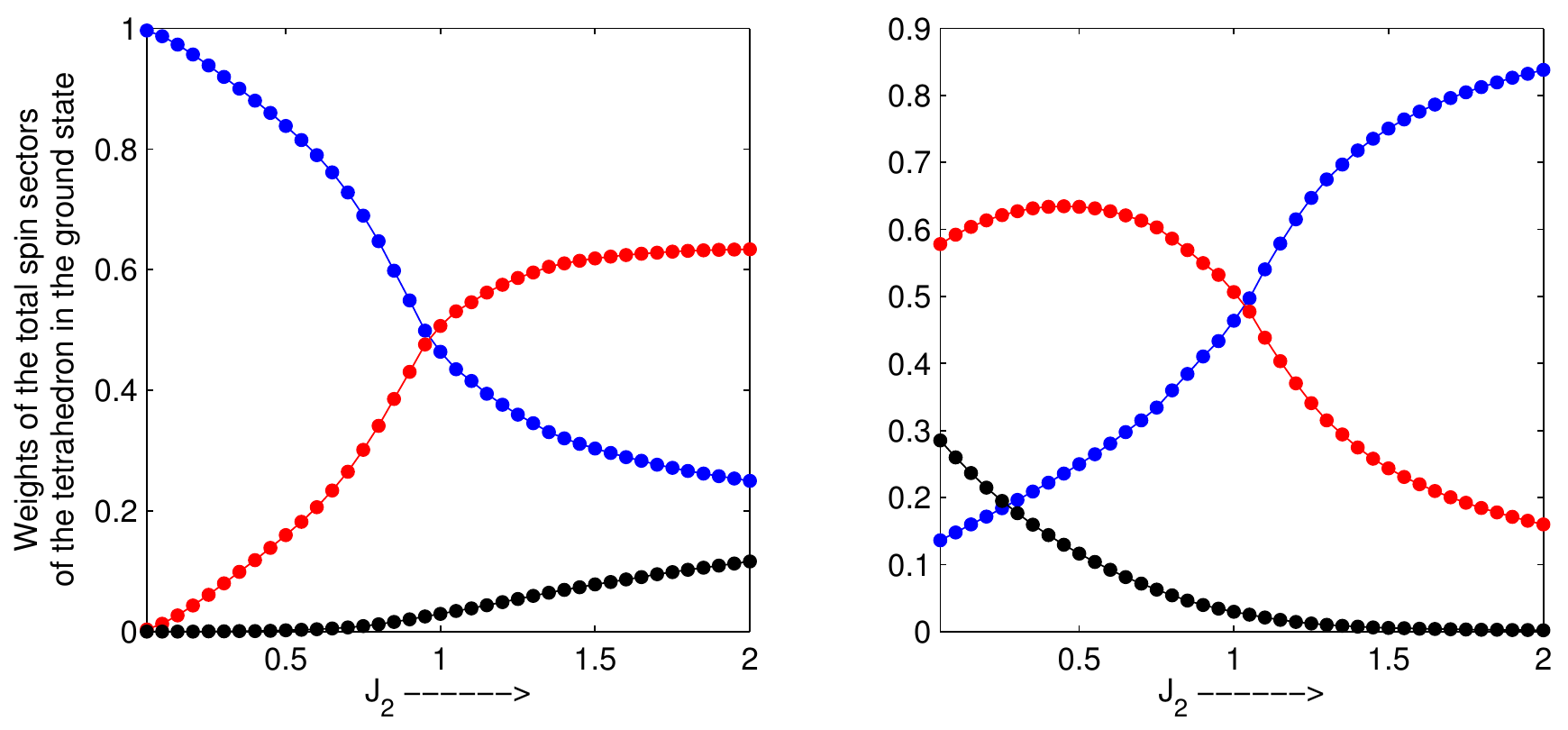}
\caption{{\small{The weights of the different total spin sectors of the tetrahedron in the Schmidt decomposed ground state
(see Eq. \ref{EqSchmidt} and the following discussion) with a tetrahedron as one of the partitions. We have shown data for the $36$ site
system and the $\mathcal{P} = -1$ symmetry sector. The blue, red, and black markers denote the weights of the singlet, triplet and quintet sectors of the tetrahedron. The figure on the
left is for a $J_1$ tetrahedron (sites $(0, 1, 4, 12)$) and the one on the right is for a $J_2$ tetrahedron (sites $(12, 22, 23, 26)$).
Weights add to $1$ as is to be expected for reduced density matrix eigenvalues.}}}
\label{Fig_tetra_spin_weights}
\end{figure*}

Fig. \ref{Fig_dim_dim_single_tetrahedron} shows the same dimer-dimer correlations as a function of $J_2$. The three pairs of nearest neighbour correlations 
in a tetrahedron have been shown explicitly.  It is clear from this figure that the system passes through several regions where the ground state is doubly degenerate, as mentioned earlier.
These distinct regions in the plot are tabulated in Table \ref{table_deg_regions} and the degeneracy is indicated using the eigenvalue of the site permutation operator $\mathcal{P}$.
It can be seen that whenever the ground state is non-degenerate the three different dimer-dimer spin correlations within a tetrahedron are
all equal. In case of degeneracy the symmetry between the dimers is broken and results in the different values 
for particular dimer-dimer correlations in the two eigenstates. We note that how exactly the broken symmetry manifests itself depends on the chosen permutation
of site indices. For our chosen permutation $\langle \phi_{0-1}~\phi_{4-12} \rangle$ and  $\langle \phi_{1-4}~\phi_{0-12} \rangle$  have the same value
in case of degeneracy and $\langle \phi_{0-4}~\phi_{1-12} \rangle$ has a different value. This might change if we had chosen another permutation from the automorphism
group to extract distinct symmetry labels for the eigenstates.\\

The figure on the right in Fig \ref{Fig_dim_dim_single_tetrahedron} shows the average of the three dimer correlations at any $J_2$. 
The effect of the degeneracy is not visible in this plot and we get a single crossing at $J_2 = 1$. The general trend of the plots is similar
for the 28 site system, though there the regions where eigenvalue is degenerate are different. Thus the 
details of the existence and locations of the several crossings might be dependent on the nature of the 
finite size system and the boundary conditions. However, if we consider the full tetrahedron (by considering an average as above) 
then at least from the point of view of dimer-dimer correlations within a single tetrahedron there is a smooth evolution from the 
state of disconnected tetrahedra at $J_2 = 0$ to the state in the vicinity of $J_2=1$.\\

We would like to understand in a bit more detail how the tetrahedral unit evolves
within the system from $J_2=0$ to $J_2 = 1$. To that end we analyse quantitatively how the tetrahedron is entangled with the rest 
of the lattice. In order to do that we make use of the property of Schmidt decomposition of a pure state. 
We know that a pure state of any quantum system (say $|\psi \rangle$) partitioned into two parts (say $A$ and $B$)
can be written as:

\beq
\vert \psi \rangle = \sum_{\lambda} \sqrt{\lambda} \vert \lambda \rangle_A \otimes \vert \lambda \rangle_B
\label{EqSchmidt}
\eeq 
Here  $\lambda$ are real and positive and the number of the terms in the sum is at most
the smaller of the Hilbert space dimensions of $A$ and $B$. $\vert \lambda \rangle_{A/B}$ are Schmidt vectors for the partition
for a given $\lambda$ and $\null_A\langle \lambda \vert \lambda^\prime \rangle_A =  \null_B\langle \lambda \vert \lambda^\prime \rangle_B = \delta_{\lambda, \lambda^{\prime}}$.
The Schmidt vectors can also be shown to be the eigenvectors of the reduced density matrix
of the partition with the eigenvalue $\lambda$.\\
\begin{figure*}[t]
\includegraphics[scale=1.0]{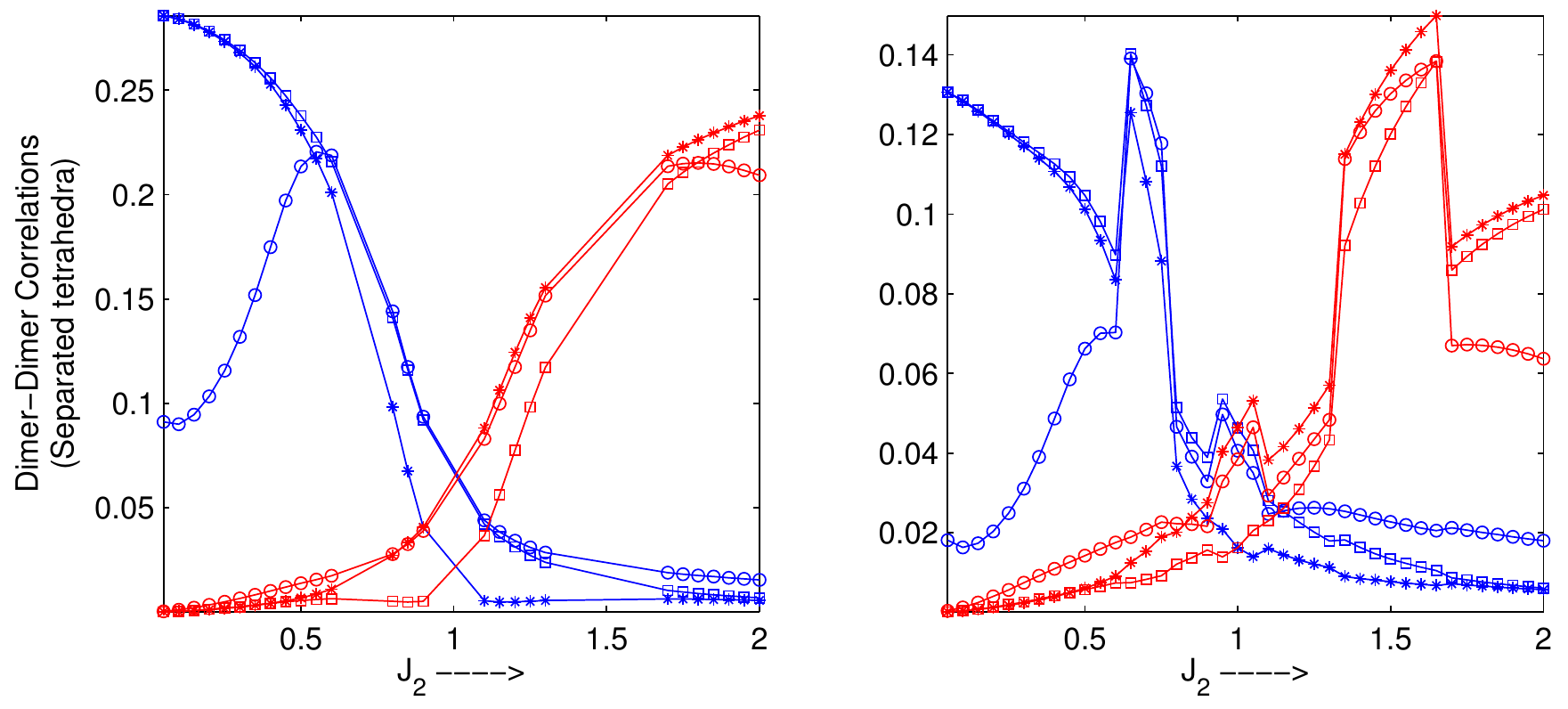}
\caption{{\small{
 Dimer-Dimer correlations ($\langle \phi_{ij}~\phi_{kl} \rangle$,  ${\phi}_{ij} \equiv  \mbf{S}_i \cdot \mbf{S}_j$ ) in the ground state across different tetrahedra as a function of $J_2$.
$(ij)$ is the $(0-12)$ for blue symbols and $(12-26)$ for the red symbols. ${(kl)}$ are $(6-14)$ (blue squares), $(2-13)$ (blue stars), $(23-31)$ (blue circles), $(13-27)$ (red stars), $(10-34) $ (red circles),
$(14-20)$ (red squares). The plot on the left is for eigenstate with $\mathcal{P} = 1$ and the right for $\mathcal{P} = -1$. All data shown in this plot is for the 36 site cluster.}}}
\label{Fig_four_point_corrs_36}
\end{figure*}

Usual studies of entanglement content of lattice models \cite{amicoetalEntRmp} involve partitioning the system of interest into two 
blocks to study the scaling of entanglement entropy as the size of one of the partitions is increased. However, Schmidt decomposition can also
be used to directly probe the part of the Hilbert space of a partition responsible for entangling it with the whole system. Here we choose
the basic tetrahedral unit as one of our partitions. Our objective is to check quantitatively which states of the tetrahedral Hilbert space
are principally responsible for its entanglement with the rest of the lattice and with what weight. Since the ground state is a singlet and
the system is bipartitioned the Schmidt vectors can be labelled using a total spin label for the tetrahedron and the sum of $\lambda$ values
corresponding to a particular total spin directly gives the weight of that total spin sector in the Schmidt decomposition.\\

Fig. \ref{Fig_tetra_spin_weights} shows the spin weights corresponding to the three total spin sectors $S_{total} = 0, 1, 2$ in the
tetrahedron in the Schmidt decomposition. We have only shown results for the ground state in the $\mathcal{P} = -1$ sector, the 
sector $\mathcal{P}=1$ is similar except for missing data points in regions where the ground state is non-degenerate with $\mathcal{P} = -1$ 
(see Table. \ref{table_deg_regions}). We see that as the system evolves from $J_2=0$ where the tetrahedra are disconnected from each other the 
weight of the triplet sector of the tetrahedral Hilbert space grows progressively at the expense of the singlet sector which
is the only relevant sector at $J_2 =0$. Notably, the weight contributed by the triplet sector of the tetrahedron near $J_2= 1$
is not a sub-dominant fraction of the total weight but is almost equal to the weight contributed by the singlet sector. Hence effective theory
formulations which rely on discarding the non-singlet states of a tetrahedron in effect discard a part of the Hilbert space which is as important 
as the singlet sector from the point of view of entanglement of the tetrahedron with the rest of the lattice.\\

In addition to the weights of the total spin sectors which depend on $\lambda$ one can also study the structure of individual Schmidt vectors
to gain detailed information about the way different possible states in a sector contribute to the total weight of that sector.
For instance, one can explicitly evaluate the weights contributed by the two possible total spin singlets 
on the tetrahedron to the overall weight contributed by the singlet sector shown in Fig. \ref{Fig_tetra_spin_weights}.
Such a computation shows that the two degenerate eigenvectors (when present) differ by the weights contributed by the two singlets simply being exchanged
with each other. In case of non-degenerate ground states the two kinds of singlets contribute equal weight. This essentially is the microscopic
reason for the structure of the dimer-dimer correlations plots shown in Fig. \ref{Fig_dim_dim_single_tetrahedron}.

The analysis of the ground state just presented, involving Schmidt decomposition with a basic unit of the lattice as a partition, is of course general 
and it can be used to gain information about the ground state structure of other spin Hamiltonians. As the size of the cluster of interest increases,
this approach enables us to extract information that may not be easily accessible using correlation functions. \\ 

\begin{figure*}[t]
\includegraphics[scale=0.80]{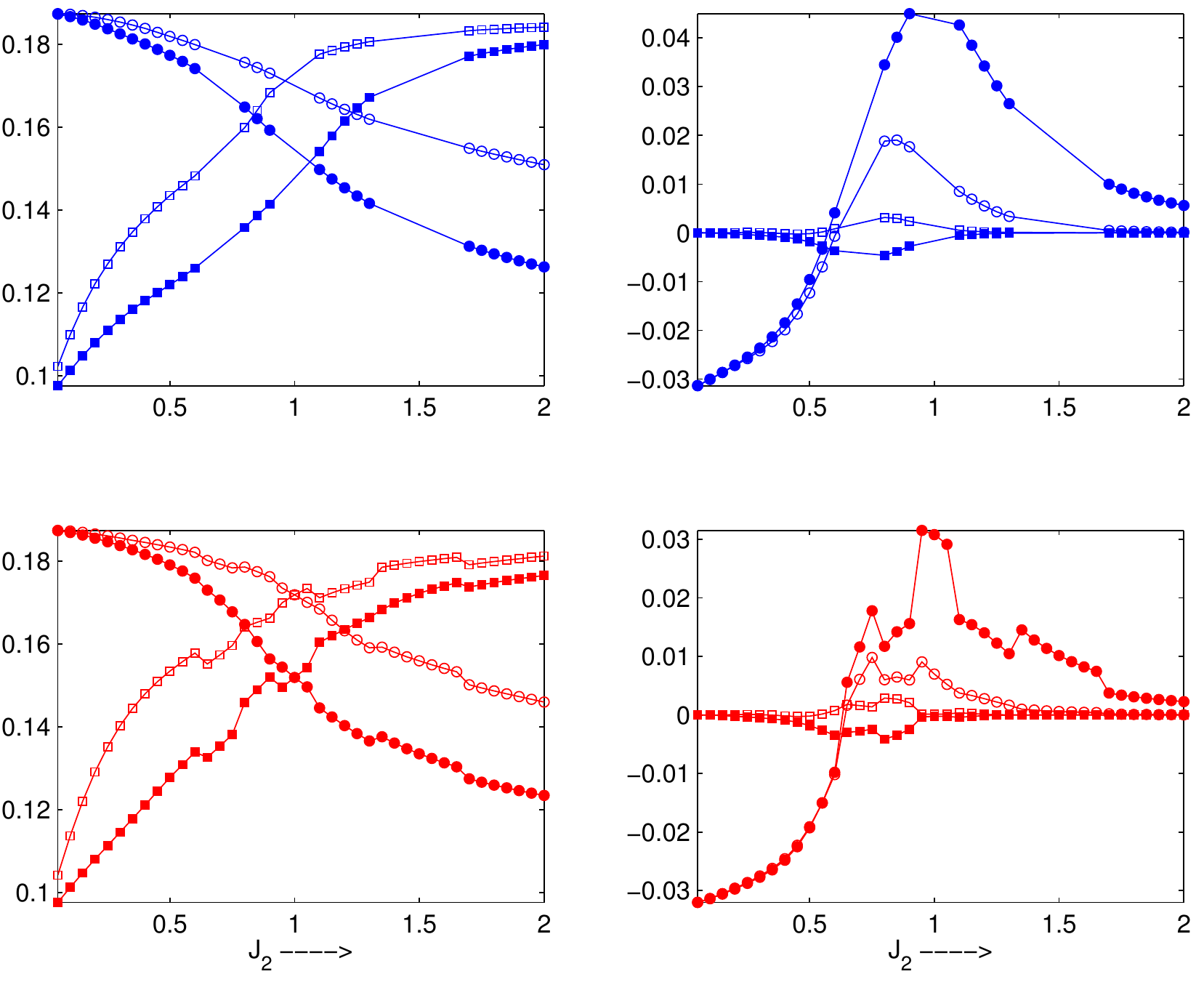}
\caption {\small{The variation of chirality correlations ($\langle \Xi_{i-j-k} \Xi_{l-m-n} \rangle,~ \Xi_{i-j-k} \equiv \mbf{S}_i \cdot (\mbf{S}_j \times \mbf{S}_k)$) with $J_2$
for the ground state of the $36$ site  cluster. The two plots on the top are for $\mathcal{P}=1$ (blue markers) and the two plots on the bottom for $\mathcal{P}=-1$ (red markers). The plots
on the left are for correlations $\langle \Xi_{i-j-k} \Xi_{i-j-k} \rangle $ ($(ijk)$ same in both triangles). $(i-j-k)$} are given by: (1-4-12) (filled circles), (12-23-26) (filled squares),
(4-8-32) (empty squares), (21-24-31) (empty circles). The two figures on the right are chirality correlations for separated triangles: $\langle \Xi_{1-4-12} \Xi_{7-10-14} \rangle $ (filled circles),
$\langle \Xi_{1-4-12} \Xi_{3-5-13} \rangle $ (empty circles), $\langle  \Xi_{1-4-12} \Xi_{21-24-31} \rangle $ (filled squares), $\langle  \Xi_{1-4-12} \Xi_{18-27-29} \rangle $ (empty squares). Note
that the number of bonds to be crossed to go from one triangle to the other is larger for the open symbols compared to the closed ones.}
\label{Fig_chirality_corrs_36}
\end{figure*}

Another set of dimer-dimer correlations shown in Table. \ref{j2onedatatable} are correlations between dimers on different tetrahedra. 
Fig. \ref{Fig_four_point_corrs_36} shows these correlations as a function of $J_2$ for the $36$ site cluster. 
The plot on the left is for the ground state with eigenvalue $+1$ of the permutation operator $\mathcal{P}$ and figure on the right are for eigenvalue $-1$.  
These correlations have been analysed to probe the possibility of order in dimer-dimer correlations which has been discussed
\cite{harris_bruder_berlinsky, tsunetsugu_2001_pyrochlore_prb,tsunetsugu_pyrochlore_jpsj, isoda_mori_pyrochlore_jpsj}
as one of the possible low temperature states at $J_2 = 1$. As can be seen by comparing Fig. \ref{Fig_dim_dim_single_tetrahedron} and Fig. \ref{Fig_four_point_corrs_36}
(dimer-dimer correlations within and between tetrahedra) there is an extended region till about  $J_2 \sim 0.75$ where these 
correlations between tetrahedra remain strong and sometimes comparable to the correlations
within a single tetrahedron, something we would expect in a state with dimer order. Following this we see a drop in correlations to low values which are smaller than 
the correlations within a tetrahedron. This is also the time when the disparity between the dimer correlations 
belonging to tetrahedra of different orientation (blue and red markers in the plot) start reducing sharply. This is something we would 
expect on general grounds as the system approaches $J_2 = 1$. In this region while there is a decrease of the correlation with distance 
as can be seen in Table \ref{j2onedatatable} and Fig. \ref{Fig_four_point_corrs_36}.
it is clearly not as rapid as in the case of the two point correlator. As a result while it is clear that
the system goes to a qualitatively different phase the currently available data 
is not sufficient to conclusively rule out some remnant weak dimer order near $J_2 = 1$.

\subsection{Chirality correlations}
\label{chirality_corrs_sec}
In section \ref{results_sec_lat_ham} we noted that the low energy degenerate ground state manifold of the elementary tetrahedral unit makes this system
a promising candidate for exploring long range chiral order. We now present the evaluation of chirality correlations in this system. 
They are denoted by  $\langle \Xi_{i-j-k} \Xi_{l-m-n} \rangle$ where,  $\Xi_{i-j-k} \equiv  \mbf{S}_i \cdot (\mbf{S}_j \times \mbf{S}_k)$ and as usual we evaluate  
the expectation value in the ground state. The evaluated chirality correlations are shown in Table \ref{j2onedatatable} for $J_2=1$. The same
correlations are shown as a function of $J_2$ in Fig. \ref{Fig_chirality_corrs_36} for 36 sites. 
The plots on the right are for correlations of the form ($\langle \Xi_{i-j-k} \Xi_{l-m-n} \rangle$) and the two triangles involved
belong to different tetrahedra. It is clear from these plots that these correlations do decay very quickly as can be seen both in the plots
and for $J_2 = 1$ in Table \ref{j2onedatatable}. While it is not feasible at these lattice sizes to directly compare predicted values 
for the chiral order parameter \cite{burnell_et_al_pyrochlore_prb, kim_and_han_pyrochlore_prb}, any presumed decay at this rate
for a few more lattice spacings would result in the order parameter being very small. \\

Variational Monte Carlo evaluations which predict chiral order usually involve imposing uniformity in the nearest neighbour two spin correlations
when choosing an ansatz for the mean field analysis. Our calculations have open boundary conditions
in the third direction and that essentially means all two spin correlations are not equivalent. We would like to check how severe is 
the effect of open boundary conditions. The plots on the left in Fig. \ref{Fig_chirality_corrs_36} are for correlations of the form
$\langle \Xi_{i-j-k} \Xi_{i-j-k} \rangle$ and these have been evaluated to get an indication of the effect of open boundary conditions in the $[111]$ direction.
The open symbols correspond to a $\Xi_{i-j-k}$ with the triangle being part of a tetrahedron which contains a site from the triangular lattice layer at the bottom
or top in Fig. \ref{Fig_pyro}. The full symbols correspond to a tetrahedron in the interior in which each site is participating in $6$ bonds. We note that the open symbols 
are consistently above the full symbols for all $J_2$. From the plot it is clear that the triangle on the boundary is more likely to behave akin to an 
"isolated" tetrahedron. We note that for a single tetrahedron $\langle \Xi_{i-j-k} \Xi_{i-j-k} \rangle$ has a value of $0.1875$ in the
ground state. This can be compared with $0.151897$ for the triad $(1-4-12)$ and $0.171852$ for the triad $(4-8-32)$ in our case for $J_2 = 1$. 
Furthermore we have checked that for a $32$ site cluster with periodic boundary conditions the value of this correlator is $0.158357$ for $J_2 = 1$.
These results empirically indicate that the effect of open boundary conditions in the third direction in our problem is probably not so severe as to 
render a possible chiral ordered state undetectable. Nevertheless, a computation using a larger lattice size can help putting this claim on a stronger footing. 
 
\section{Summary and Discussion}

This study hopes to contribute to the existing understanding of the spin-1/2 antiferromagnet on the pyrochlore lattice
from an exact diagonalisation perspective.\\
From the point of view of the nature of the low lying spectrum and the two spin correlations the system 
shares common features of quantum frustrated magnetism also found in other systems. The short length scale of spin 
correlations is one such feature for $J_2 = 1$ and our study confirms that, though at this lattice size we cannot 
formally extract a correlation length. We have also shown the presence of several low lying singlet excitations below 
the magnetic gap.

We investigated the dimer-dimer and scalar chirality correlations in some detail in Sec. \ref{gscorrs}. Briefly, we can
say that while not being entirely conclusive, ED data for the system we have studied do not strengthen
the case for the existence of these orders at $J_2 = 1$. 

In case of the dimer-dimer correlations several of the earlier predictions involve a starting point where the
Hilbert space of the tetrahedral unit is truncated to the two degenerate singlets. We showed using an analysis
of the Schmidt coefficients that result from using the tetrahedron as one of the partitions that the triplet
sector of the tetrahedron contributes almost an equal weight near $J_2=1$. This fact in conjunction with the
observation that dimer correlations are small and decaying (albeit slowly) leads us to suspect that dimer order even
if it is present is likely to be very weak. For the chirality correlations we find that the decay is more rapid 
than in the case of dimer-dimer correlations and the value of correlations is quite small. 
We note here, as also pointed out in the discussion regarding Fig. \ref{Fig_dim_dim_single_tetrahedron} in Sec. \ref{dimer_dimer_corrs},
that the seemingly abrupt changes in correlations in Figs. \ref{Fig_dim_dim_single_tetrahedron}, \ref{Fig_four_point_corrs_36}, 
\ref{Fig_chirality_corrs_36} have their origins in the degeneracy of the spectrum for this specific cluster. We have 
verified that for a smaller cluster of $32$ sites with fully periodic boundary conditions (where the data does not have such abrupt changes)
all the statements about the lack of the studied orders still hold. 

In order to make the claims of the lack of the above orders stronger it is important to evaluate the same correlations for
a larger symmetric cluster which can give access to more separation between the dimers and triangles involved
in the evaluated correlations. One clear way of doing this would be to add another Kagome layer of $12$ sites in 
Fig.\ref{Fig_111_pyro} and then coupling it to the triangular lattice layer on the other side using periodic boundary conditions. 
This would result in a fully periodic cluster of 48 spins. This lattice size, though accessible with current technology \cite{kagome_ED_48_sites} 
is considerably more involved and is beyond the scope of this paper. We can not also in this work comment on dynamical 
correlations of this model which might have a bearing on the detailed description of the nature of the
possible $T\rightarrow 0 $ liquid state. These analyses will help determine the course ahead to a more complete understanding
of the properties of this model.
\appendix
\section{}
\label{appendix_permutation}
As mentioned in the text the implementation of the conservation of $S_z^{total}$ and spin inversion symmetry
for exact diagonalisation is not sufficient to get a non-degenerate ground state in each symmetry sector. 
In order to analyse orthogonal eigenvectors with distinct symmetry labels 
we use the following permutation of site indices, which is an element of the automorphism
group of the graph and results in a symmetry of the Hamiltonian, for any $J_2$. 
\par
$(0,5,8,10) (1,3,9,7) (2,11,6,4) (12,13,15,14)\\ 
\hspace{0.75cm} (16,22,24,18) (17,26,25,21) (19,20,23,27) \\
(28,30,31,29) (32,34,35,33)$\\
\par
Throughout the text this site permutation is referred to as $\mathcal{P}$.  
Here  $(a, b, c, d)$ denotes a cycle representing the site permutation $(a\rightarrow b \rightarrow c \rightarrow d \rightarrow a)$.
The above permutation is an element of order $4$ of the automorphism group for the graph representing the considered lattice. 
The evaluated eigenvectors belong either to eigenvalues $1$ or $-1$ as indicated in the text. We note that this is one of several possible
elements of the symmetry group of the Hamiltonian. It would be perfectly legitimate to choose another permutation, it
would merely result in a different linear combination of the eigenvectors. Information regarding the automorphism group
has been extracted using the GAP, GRAPE and NAUTY software packages.

\acknowledgments
VRC thanks the Centre for Development of Advanced Computing, Pune, India for computational
resources through the project NumericalStudy-Magnets-PR and Sylvain Capponi for e-mail communication with
helpful pointers regarding the usage of the GAP software package.
\bibliography{spin_half_pyrochlore_references}

\begin{thebibliography}{25}%
\makeatletter
\providecommand \@ifxundefined [1]{%
 \@ifx{#1\undefined}
}%
\providecommand \@ifnum [1]{%
 \ifnum #1\expandafter \@firstoftwo
 \else \expandafter \@secondoftwo
 \fi
}%
\providecommand \@ifx [1]{%
 \ifx #1\expandafter \@firstoftwo
 \else \expandafter \@secondoftwo
 \fi
}%
\providecommand \natexlab [1]{#1}%
\providecommand \enquote  [1]{``#1''}%
\providecommand \bibnamefont  [1]{#1}%
\providecommand \bibfnamefont [1]{#1}%
\providecommand \citenamefont [1]{#1}%
\providecommand \href@noop [0]{\@secondoftwo}%
\providecommand \href [0]{\begingroup \@sanitize@url \@href}%
\providecommand \@href[1]{\@@startlink{#1}\@@href}%
\providecommand \@@href[1]{\endgroup#1\@@endlink}%
\providecommand \@sanitize@url [0]{\catcode `\\12\catcode `\$12\catcode
  `\&12\catcode `\#12\catcode `\^12\catcode `\_12\catcode `\%12\relax}%
\providecommand \@@startlink[1]{}%
\providecommand \@@endlink[0]{}%
\providecommand \url  [0]{\begingroup\@sanitize@url \@url }%
\providecommand \@url [1]{\endgroup\@href {#1}{\urlprefix }}%
\providecommand \urlprefix  [0]{URL }%
\providecommand \Eprint [0]{\href }%
\providecommand \doibase [0]{http://dx.doi.org/}%
\providecommand \selectlanguage [0]{\@gobble}%
\providecommand \bibinfo  [0]{\@secondoftwo}%
\providecommand \bibfield  [0]{\@secondoftwo}%
\providecommand \translation [1]{[#1]}%
\providecommand \BibitemOpen [0]{}%
\providecommand \bibitemStop [0]{}%
\providecommand \bibitemNoStop [0]{.\EOS\space}%
\providecommand \EOS [0]{\spacefactor3000\relax}%
\providecommand \BibitemShut  [1]{\csname bibitem#1\endcsname}%
\let\auto@bib@innerbib\@empty
\bibitem [{\citenamefont {Wannier}(1950)}]{wannier_triangular_lattice_1950}%
  \BibitemOpen
  \bibfield  {author} {\bibinfo {author} {\bibfnamefont {G.~H.}\ \bibnamefont
  {Wannier}},\ }\href {\doibase 10.1103/PhysRev.79.357} {\bibfield  {journal}
  {\bibinfo  {journal} {Phys. Rev.}\ }\textbf {\bibinfo {volume} {79}},\
  \bibinfo {pages} {357} (\bibinfo {year} {1950})}\BibitemShut {NoStop}%
\bibitem [{\citenamefont {Anderson}(1956)}]{anderson_Ising_pyrochlore_1956}%
  \BibitemOpen
  \bibfield  {author} {\bibinfo {author} {\bibfnamefont {P.~W.}\ \bibnamefont
  {Anderson}},\ }\href {\doibase 10.1103/PhysRev.102.1008} {\bibfield
  {journal} {\bibinfo  {journal} {Phys. Rev.}\ }\textbf {\bibinfo {volume}
  {102}},\ \bibinfo {pages} {1008} (\bibinfo {year} {1956})}\BibitemShut
  {NoStop}%
\bibitem [{\citenamefont {Gardner}\ \emph {et~al.}(2010)\citenamefont
  {Gardner}, \citenamefont {Gingras},\ and\ \citenamefont
  {Greedan}}]{gardner_gingras_greedan_rmp}%
  \BibitemOpen
  \bibfield  {author} {\bibinfo {author} {\bibfnamefont {J.~S.}\ \bibnamefont
  {Gardner}}, \bibinfo {author} {\bibfnamefont {M.~J.~P.}\ \bibnamefont
  {Gingras}}, \ and\ \bibinfo {author} {\bibfnamefont {J.~E.}\ \bibnamefont
  {Greedan}},\ }\href {\doibase 10.1103/RevModPhys.82.53} {\bibfield  {journal}
  {\bibinfo  {journal} {Rev. Mod. Phys.}\ }\textbf {\bibinfo {volume} {82}},\
  \bibinfo {pages} {53} (\bibinfo {year} {2010})}\BibitemShut {NoStop}%
\bibitem [{\citenamefont {Kimura}\ \emph {et~al.}(2014)\citenamefont {Kimura},
  \citenamefont {Nakatsuji},\ and\ \citenamefont
  {Kimura}}]{breathing_pyrochlore_kimura_et_al_2014}%
  \BibitemOpen
  \bibfield  {author} {\bibinfo {author} {\bibfnamefont {K.}~\bibnamefont
  {Kimura}}, \bibinfo {author} {\bibfnamefont {S.}~\bibnamefont {Nakatsuji}}, \
  and\ \bibinfo {author} {\bibfnamefont {T.}~\bibnamefont {Kimura}},\ }\href
  {\doibase 10.1103/PhysRevB.90.060414} {\bibfield  {journal} {\bibinfo
  {journal} {Phys. Rev. B}\ }\textbf {\bibinfo {volume} {90}},\ \bibinfo
  {pages} {060414} (\bibinfo {year} {2014})}\BibitemShut {NoStop}%
\bibitem [{\citenamefont {Clark}\ \emph {et~al.}(2014)\citenamefont {Clark},
  \citenamefont {Nilsen}, \citenamefont {Kermarrec}, \citenamefont {Ehlers},
  \citenamefont {Knight}, \citenamefont {Harrison}, \citenamefont {Attfield},\
  and\ \citenamefont {Gaulin}}]{molybdate_pyrochlore_gaulin_et_al_2014}%
  \BibitemOpen
  \bibfield  {author} {\bibinfo {author} {\bibfnamefont {L.}~\bibnamefont
  {Clark}}, \bibinfo {author} {\bibfnamefont {G.~J.}\ \bibnamefont {Nilsen}},
  \bibinfo {author} {\bibfnamefont {E.}~\bibnamefont {Kermarrec}}, \bibinfo
  {author} {\bibfnamefont {G.}~\bibnamefont {Ehlers}}, \bibinfo {author}
  {\bibfnamefont {K.~S.}\ \bibnamefont {Knight}}, \bibinfo {author}
  {\bibfnamefont {A.}~\bibnamefont {Harrison}}, \bibinfo {author}
  {\bibfnamefont {J.~P.}\ \bibnamefont {Attfield}}, \ and\ \bibinfo {author}
  {\bibfnamefont {B.~D.}\ \bibnamefont {Gaulin}},\ }\href {\doibase
  10.1103/PhysRevLett.113.117201} {\bibfield  {journal} {\bibinfo  {journal}
  {Phys. Rev. Lett.}\ }\textbf {\bibinfo {volume} {113}},\ \bibinfo {pages}
  {117201} (\bibinfo {year} {2014})}\BibitemShut {NoStop}%
\bibitem [{\citenamefont {Iqbal}\ \emph {et~al.}(2017)\citenamefont {Iqbal},
  \citenamefont {Müller}, \citenamefont {Riedl}, \citenamefont {Reuther},
  \citenamefont {Rachel}, \citenamefont {Valentí}, \citenamefont {Gingras},
  \citenamefont {Thomale},\ and\ \citenamefont
  {Jeschke}}]{Yiqbaletal_gearwheel_spin_liquid}%
  \BibitemOpen
  \bibfield  {author} {\bibinfo {author} {\bibfnamefont {Y.}~\bibnamefont
  {Iqbal}}, \bibinfo {author} {\bibfnamefont {T.}~\bibnamefont {Müller}},
  \bibinfo {author} {\bibfnamefont {K.}~\bibnamefont {Riedl}}, \bibinfo
  {author} {\bibfnamefont {J.}~\bibnamefont {Reuther}}, \bibinfo {author}
  {\bibfnamefont {S.}~\bibnamefont {Rachel}}, \bibinfo {author} {\bibfnamefont
  {R.}~\bibnamefont {Valentí}}, \bibinfo {author} {\bibfnamefont {M.~J.~P.}\
  \bibnamefont {Gingras}}, \bibinfo {author} {\bibfnamefont {R.}~\bibnamefont
  {Thomale}}, \ and\ \bibinfo {author} {\bibfnamefont {H.~O.}\ \bibnamefont
  {Jeschke}},\ }\href@noop {} {} (\bibinfo {year} {2017}),\ \Eprint
  {http://arxiv.org/abs/arXiv:1705.05291} {arXiv:1705.05291} \BibitemShut
  {NoStop}%
\bibitem [{\citenamefont {Harris}\ \emph {et~al.}(1991)\citenamefont {Harris},
  \citenamefont {Berlinsky},\ and\ \citenamefont
  {Bruder}}]{harris_bruder_berlinsky}%
  \BibitemOpen
  \bibfield  {author} {\bibinfo {author} {\bibfnamefont {A.~B.}\ \bibnamefont
  {Harris}}, \bibinfo {author} {\bibfnamefont {A.~J.}\ \bibnamefont
  {Berlinsky}}, \ and\ \bibinfo {author} {\bibfnamefont {C.}~\bibnamefont
  {Bruder}},\ }\href {http://aip.scitation.org/doi/10.1063/1.348098} {\bibfield
   {journal} {\bibinfo  {journal} {Journal of Applied Physics}\ }\textbf
  {\bibinfo {volume} {69}},\ \bibinfo {pages} {5200} (\bibinfo {year}
  {1991})}\BibitemShut {NoStop}%
\bibitem [{\citenamefont
  {Tsunetsugu}(2001{\natexlab{a}})}]{tsunetsugu_2001_pyrochlore_prb}%
  \BibitemOpen
  \bibfield  {author} {\bibinfo {author} {\bibfnamefont {H.}~\bibnamefont
  {Tsunetsugu}},\ }\href {\doibase 10.1103/PhysRevB.65.024415} {\bibfield
  {journal} {\bibinfo  {journal} {Phys. Rev. B}\ }\textbf {\bibinfo {volume}
  {65}},\ \bibinfo {pages} {024415} (\bibinfo {year}
  {2001}{\natexlab{a}})}\BibitemShut {NoStop}%
\bibitem [{\citenamefont
  {Tsunetsugu}(2001{\natexlab{b}})}]{tsunetsugu_pyrochlore_jpsj}%
  \BibitemOpen
  \bibfield  {author} {\bibinfo {author} {\bibfnamefont {H.}~\bibnamefont
  {Tsunetsugu}},\ }\href {\doibase 10.1143/JPSJ.70.640} {\bibfield  {journal}
  {\bibinfo  {journal} {Journal of the Physical Society of Japan}\ }\textbf
  {\bibinfo {volume} {70}},\ \bibinfo {pages} {640} (\bibinfo {year}
  {2001}{\natexlab{b}})}\BibitemShut {NoStop}%
\bibitem [{\citenamefont {Isoda}\ and\ \citenamefont
  {Mori}(1998)}]{isoda_mori_pyrochlore_jpsj}%
  \BibitemOpen
  \bibfield  {author} {\bibinfo {author} {\bibfnamefont {M.}~\bibnamefont
  {Isoda}}\ and\ \bibinfo {author} {\bibfnamefont {S.}~\bibnamefont {Mori}},\
  }\href {\doibase 10.1143/JPSJ.67.4022} {\bibfield  {journal} {\bibinfo
  {journal} {Journal of the Physical Society of Japan}\ }\textbf {\bibinfo
  {volume} {67}},\ \bibinfo {pages} {4022} (\bibinfo {year}
  {1998})}\BibitemShut {NoStop}%
\bibitem [{\citenamefont {Tsunetsugu}(2017)}]{tsunetsugu_2017_pyrochlore_ptep}%
  \BibitemOpen
  \bibfield  {author} {\bibinfo {author} {\bibfnamefont {H.}~\bibnamefont
  {Tsunetsugu}},\ }\href@noop {} {\bibfield  {journal} {\bibinfo  {journal}
  {Progress of Theoretical and Experimental Physics}\ }\textbf {\bibinfo
  {volume} {2017}},\ \bibinfo {pages} {033I01} (\bibinfo {year}
  {2017})}\BibitemShut {NoStop}%
\bibitem [{\citenamefont {Berg}\ \emph {et~al.}(2003)\citenamefont {Berg},
  \citenamefont {Altman},\ and\ \citenamefont
  {Auerbach}}]{berg_altman_auerbach_prl}%
  \BibitemOpen
  \bibfield  {author} {\bibinfo {author} {\bibfnamefont {E.}~\bibnamefont
  {Berg}}, \bibinfo {author} {\bibfnamefont {E.}~\bibnamefont {Altman}}, \ and\
  \bibinfo {author} {\bibfnamefont {A.}~\bibnamefont {Auerbach}},\ }\href
  {\doibase 10.1103/PhysRevLett.90.147204} {\bibfield  {journal} {\bibinfo
  {journal} {Phys. Rev. Lett.}\ }\textbf {\bibinfo {volume} {90}},\ \bibinfo
  {pages} {147204} (\bibinfo {year} {2003})}\BibitemShut {NoStop}%
\bibitem [{\citenamefont {Canals}\ and\ \citenamefont
  {Lacroix}(1998)}]{canals_and_lacroix_pyrochlore_prl}%
  \BibitemOpen
  \bibfield  {author} {\bibinfo {author} {\bibfnamefont {B.}~\bibnamefont
  {Canals}}\ and\ \bibinfo {author} {\bibfnamefont {C.}~\bibnamefont
  {Lacroix}},\ }\href {\doibase 10.1103/PhysRevLett.80.2933} {\bibfield
  {journal} {\bibinfo  {journal} {Phys. Rev. Lett.}\ }\textbf {\bibinfo
  {volume} {80}},\ \bibinfo {pages} {2933} (\bibinfo {year}
  {1998})}\BibitemShut {NoStop}%
\bibitem [{\citenamefont {Moessner}\ \emph {et~al.}(2006)\citenamefont
  {Moessner}, \citenamefont {Sondhi},\ and\ \citenamefont
  {Goerbig}}]{goerbig_moessner_sondhi_qdm_prb}%
  \BibitemOpen
  \bibfield  {author} {\bibinfo {author} {\bibfnamefont {R.}~\bibnamefont
  {Moessner}}, \bibinfo {author} {\bibfnamefont {S.~L.}\ \bibnamefont
  {Sondhi}}, \ and\ \bibinfo {author} {\bibfnamefont {M.~O.}\ \bibnamefont
  {Goerbig}},\ }\href {\doibase 10.1103/PhysRevB.73.094430} {\bibfield
  {journal} {\bibinfo  {journal} {Phys. Rev. B}\ }\textbf {\bibinfo {volume}
  {73}},\ \bibinfo {pages} {094430} (\bibinfo {year} {2006})}\BibitemShut
  {NoStop}%
\bibitem [{\citenamefont {Tchernyshyov}\ \emph {et~al.}(2006)\citenamefont
  {Tchernyshyov}, \citenamefont {Moessner},\ and\ \citenamefont
  {Sondhi}}]{tchernyshyov_moessner_sondhi_large_N_epl}%
  \BibitemOpen
  \bibfield  {author} {\bibinfo {author} {\bibfnamefont {O.}~\bibnamefont
  {Tchernyshyov}}, \bibinfo {author} {\bibfnamefont {R.}~\bibnamefont
  {Moessner}}, \ and\ \bibinfo {author} {\bibfnamefont {S.~L.}\ \bibnamefont
  {Sondhi}},\ }\href {http://stacks.iop.org/0295-5075/73/i=2/a=278} {\bibfield
  {journal} {\bibinfo  {journal} {EPL (Europhysics Letters)}\ }\textbf
  {\bibinfo {volume} {73}},\ \bibinfo {pages} {278} (\bibinfo {year}
  {2006})}\BibitemShut {NoStop}%
\bibitem [{\citenamefont {Burnell}\ \emph {et~al.}(2009)\citenamefont
  {Burnell}, \citenamefont {Chakravarty},\ and\ \citenamefont
  {Sondhi}}]{burnell_et_al_pyrochlore_prb}%
  \BibitemOpen
  \bibfield  {author} {\bibinfo {author} {\bibfnamefont {F.~J.}\ \bibnamefont
  {Burnell}}, \bibinfo {author} {\bibfnamefont {S.}~\bibnamefont
  {Chakravarty}}, \ and\ \bibinfo {author} {\bibfnamefont {S.~L.}\ \bibnamefont
  {Sondhi}},\ }\href {\doibase 10.1103/PhysRevB.79.144432} {\bibfield
  {journal} {\bibinfo  {journal} {Phys. Rev. B}\ }\textbf {\bibinfo {volume}
  {79}},\ \bibinfo {pages} {144432} (\bibinfo {year} {2009})}\BibitemShut
  {NoStop}%
\bibitem [{\citenamefont {Kim}\ and\ \citenamefont
  {Han}(2008)}]{kim_and_han_pyrochlore_prb}%
  \BibitemOpen
  \bibfield  {author} {\bibinfo {author} {\bibfnamefont {J.~H.}\ \bibnamefont
  {Kim}}\ and\ \bibinfo {author} {\bibfnamefont {J.~H.}\ \bibnamefont {Han}},\
  }\href {\doibase 10.1103/PhysRevB.78.180410} {\bibfield  {journal} {\bibinfo
  {journal} {Phys. Rev. B}\ }\textbf {\bibinfo {volume} {78}},\ \bibinfo
  {pages} {180410} (\bibinfo {year} {2008})}\BibitemShut {NoStop}%
\bibitem [{\citenamefont {Huang}\ \emph {et~al.}(2016)\citenamefont {Huang},
  \citenamefont {Chen}, \citenamefont {Deng}, \citenamefont {Prokof'ev},\ and\
  \citenamefont {Svistunov}}]{svistunov_et_al_dmc_prl}%
  \BibitemOpen
  \bibfield  {author} {\bibinfo {author} {\bibfnamefont {Y.}~\bibnamefont
  {Huang}}, \bibinfo {author} {\bibfnamefont {K.}~\bibnamefont {Chen}},
  \bibinfo {author} {\bibfnamefont {Y.}~\bibnamefont {Deng}}, \bibinfo {author}
  {\bibfnamefont {N.}~\bibnamefont {Prokof'ev}}, \ and\ \bibinfo {author}
  {\bibfnamefont {B.}~\bibnamefont {Svistunov}},\ }\href {\doibase
  10.1103/PhysRevLett.116.177203} {\bibfield  {journal} {\bibinfo  {journal}
  {Phys. Rev. Lett.}\ }\textbf {\bibinfo {volume} {116}},\ \bibinfo {pages}
  {177203} (\bibinfo {year} {2016})}\BibitemShut {NoStop}%
\bibitem [{\citenamefont {Cullum}\ and\ \citenamefont
  {Willoughby}(1985)}]{cullum_and_willoughby_lanczos_1985}%
  \BibitemOpen
  \bibfield  {author} {\bibinfo {author} {\bibfnamefont {J.}~\bibnamefont
  {Cullum}}\ and\ \bibinfo {author} {\bibfnamefont {R.}~\bibnamefont
  {Willoughby}},\ }\href@noop {} {\emph {\bibinfo {title} {Lanczos Algorithms
  for Large Symmetric Eigenvalue Computations}}}\ (\bibinfo  {publisher}
  {Birkh\"{a}user},\ \bibinfo {year} {1985})\BibitemShut {NoStop}%
\bibitem [{\citenamefont {Wen}\ \emph {et~al.}(1989)\citenamefont {Wen},
  \citenamefont {Wilczek},\ and\ \citenamefont {Zee}}]{wenwilczekzeechiral}%
  \BibitemOpen
  \bibfield  {author} {\bibinfo {author} {\bibfnamefont {X.~G.}\ \bibnamefont
  {Wen}}, \bibinfo {author} {\bibfnamefont {F.}~\bibnamefont {Wilczek}}, \ and\
  \bibinfo {author} {\bibfnamefont {A.}~\bibnamefont {Zee}},\ }\href {\doibase
  10.1103/PhysRevB.39.11413} {\bibfield  {journal} {\bibinfo  {journal} {Phys.
  Rev. B}\ }\textbf {\bibinfo {volume} {39}},\ \bibinfo {pages} {11413}
  (\bibinfo {year} {1989})}\BibitemShut {NoStop}%
\bibitem [{\citenamefont {Waldtmann}\ \emph {et~al.}(1998)\citenamefont
  {Waldtmann}, \citenamefont {Everts}, \citenamefont {Bernu}, \citenamefont
  {Lhuillier}, \citenamefont {Sindzingre}, \citenamefont {Lecheminant},\ and\
  \citenamefont {Pierre}}]{waldtmann_et_al_kagome_1998}%
  \BibitemOpen
  \bibfield  {author} {\bibinfo {author} {\bibfnamefont {C.}~\bibnamefont
  {Waldtmann}}, \bibinfo {author} {\bibfnamefont {H.-U.}\ \bibnamefont
  {Everts}}, \bibinfo {author} {\bibfnamefont {B.}~\bibnamefont {Bernu}},
  \bibinfo {author} {\bibfnamefont {C.}~\bibnamefont {Lhuillier}}, \bibinfo
  {author} {\bibfnamefont {P.}~\bibnamefont {Sindzingre}}, \bibinfo {author}
  {\bibfnamefont {P.}~\bibnamefont {Lecheminant}}, \ and\ \bibinfo {author}
  {\bibfnamefont {L.}~\bibnamefont {Pierre}},\ }\href {\doibase
  10.1007/s100510050274} {\bibfield  {journal} {\bibinfo  {journal} {The
  European Physical Journal B - Condensed Matter and Complex Systems}\ }\textbf
  {\bibinfo {volume} {2}},\ \bibinfo {pages} {501} (\bibinfo {year}
  {1998})}\BibitemShut {NoStop}%
\bibitem [{\citenamefont {Palmer}\ and\ \citenamefont
  {Chalker}(2001)}]{palmer_and_chalker_checkerboard_ED_2001}%
  \BibitemOpen
  \bibfield  {author} {\bibinfo {author} {\bibfnamefont {S.~E.}\ \bibnamefont
  {Palmer}}\ and\ \bibinfo {author} {\bibfnamefont {J.~T.}\ \bibnamefont
  {Chalker}},\ }\href {\doibase 10.1103/PhysRevB.64.094412} {\bibfield
  {journal} {\bibinfo  {journal} {Phys. Rev. B}\ }\textbf {\bibinfo {volume}
  {64}},\ \bibinfo {pages} {094412} (\bibinfo {year} {2001})}\BibitemShut
  {NoStop}%
\bibitem [{\citenamefont {Fouet}\ \emph {et~al.}(2003)\citenamefont {Fouet},
  \citenamefont {Mambrini}, \citenamefont {Sindzingre},\ and\ \citenamefont
  {Lhuillier}}]{fouet_et_al_checkerboard_2003}%
  \BibitemOpen
  \bibfield  {author} {\bibinfo {author} {\bibfnamefont {J.-B.}\ \bibnamefont
  {Fouet}}, \bibinfo {author} {\bibfnamefont {M.}~\bibnamefont {Mambrini}},
  \bibinfo {author} {\bibfnamefont {P.}~\bibnamefont {Sindzingre}}, \ and\
  \bibinfo {author} {\bibfnamefont {C.}~\bibnamefont {Lhuillier}},\ }\href
  {\doibase 10.1103/PhysRevB.67.054411} {\bibfield  {journal} {\bibinfo
  {journal} {Phys. Rev. B}\ }\textbf {\bibinfo {volume} {67}},\ \bibinfo
  {pages} {054411} (\bibinfo {year} {2003})}\BibitemShut {NoStop}%
\bibitem [{\citenamefont {Amico}\ \emph {et~al.}(2008)\citenamefont {Amico},
  \citenamefont {Fazio}, \citenamefont {Osterloh},\ and\ \citenamefont
  {Vedral}}]{amicoetalEntRmp}%
  \BibitemOpen
  \bibfield  {author} {\bibinfo {author} {\bibfnamefont {L.}~\bibnamefont
  {Amico}}, \bibinfo {author} {\bibfnamefont {R.}~\bibnamefont {Fazio}},
  \bibinfo {author} {\bibfnamefont {A.}~\bibnamefont {Osterloh}}, \ and\
  \bibinfo {author} {\bibfnamefont {V.}~\bibnamefont {Vedral}},\ }\href
  {\doibase 10.1103/RevModPhys.80.517} {\bibfield  {journal} {\bibinfo
  {journal} {Rev. Mod. Phys.}\ }\textbf {\bibinfo {volume} {80}},\ \bibinfo
  {pages} {517} (\bibinfo {year} {2008})}\BibitemShut {NoStop}%
\bibitem [{\citenamefont {L\"{a}uchli}\ \emph {et~al.}(2016)\citenamefont
  {L\"{a}uchli}, \citenamefont {Sudan},\ and\ \citenamefont
  {Moessner}}]{kagome_ED_48_sites}%
  \BibitemOpen
  \bibfield  {author} {\bibinfo {author} {\bibfnamefont {A.~M.}\ \bibnamefont
  {L\"{a}uchli}}, \bibinfo {author} {\bibfnamefont {J.}~\bibnamefont {Sudan}},
  \ and\ \bibinfo {author} {\bibfnamefont {R.}~\bibnamefont {Moessner}},\
  }\href@noop {} {\enquote {\bibinfo {title} {The $s=1/2$ kagome heisenberg
  antiferromagnet revisited},}\ } (\bibinfo {year} {2016}),\ \Eprint
  {http://arxiv.org/abs/arXiv:1611.06990} {arXiv:1611.06990} \BibitemShut
  {NoStop}%
\end{thebibliography}%

\end{document}